
%
%
%
%
\magnification=1200
\magnification=1200

\countdef\refno=30
\refno=0
\countdef\sectno=31
\sectno=0
\countdef\chapno=32
\chapno=0
\def\ref{\advance \refno by 1 \ifnum\refno<10 \item{ [\the\refno ]} \else
\item{[\the\refno ]} \fi}
\outer\def\section#1#2\par
           {
           \vskip0pt plus .3\vsize\penalty-250\vskip 0pt plus-.3\vsize
           \bigskip\vskip\parskip
           \noindent\leftline{\rlap{\bf #1}
           \hskip 17pt{\bf #2}}
            \nobreak\smallskip}
\def\abstract{\vfill\eject{\bf Abstract}\smallskip}

\def\hangit#1#2\par{\setbox1=\hbox{#1\enspace}
\hangindent\wd1\hangafter=0\noindent\hskip-\wd1
\hbox{#1\enspace}\ignorespaces#2\par}

\def\Zint{{Z \kern -.45 em Z}}
\def\complex{{\kern .1em {\raise .47ex \hbox {$\scriptscriptstyle |$}}
\kern -.4em {\rm C}}}
\def\real{{\vrule height 1.6ex width 0.05em depth 0ex
\kern -0.06em {\rm R}}}
\def\rational{{\kern .1em {\raise .47ex \hbox{$\scripscriptstyle |$}}
\kern -.35em {\rm Q}}}
\def\natural{{\vrule height 1.6ex width .05em depth 0ex \kern -.35em {\rm N}}}
\def\vide{{{\rm O} \kern -0.7em /}}

\parskip 0.5truecm
\baselineskip=12pt
\catcode`\^^?=9
\newskip\zmineskip \zmineskip=0pt plus0pt minus0pt
\mathchardef\mineMM=20000
\newinsert\footins
\def\footnote#1{\let\minesf\empty 
  \ifhmode\edef\minesf{\spacefactor\the\spacefactor}\/\fi
   #1\minesf\vfootnote{#1}}
\def\vfootnote#1{\insert\footins\bgroup
  \interlinepenalty\interfootnotelinepenalty
  \splittopskip\ht\strutbox 
  \splitmaxdepth\dp\strutbox \floatingpenalty\mineMM
  \leftskip\zmineskip \rightskip\zmineskip
\spaceskip\zmineskip \xspaceskip\zmineskip
 \item{#1}\footstrut\futurelet\next\fominet}
\def\fominet{\ifcat\bgroup\noexpand\next \let\next\fmineminet
  \else\let\next\fminet\fi \next}
\def\fmineminet{\bgroup\aftergroup\minefoot\let\next}
\def\fminet#1{#1\minefoot}
\def\minefoot{\strut\egroup}
\def\footstrut{\vbox to\splittopskip{}}
\skip\footins=\bigskipamount 
\count\footins=1000 
\dimen\footins=8in 
\nopagenumbers
$$\vbox{
\vskip 4.5truecm}$$
\centerline{\bf THE BESS MODEL AT $e^+ e^-$ COLLIDERS$\;^{*}$}
\vskip 2.5truecm
\centerline {R. Casalbuoni$\;^{a,b)}$,
P. Chiappetta $\;^{c)}$,
A. Deandrea$\;^{d)}$,
S. De Curtis$\;^{b)}$,}
\smallskip
\centerline {D. Dominici$\;^{a,b)}$
and R. Gatto$\;^{d)}$}
\vskip 5truecm
\centerline{UGVA-DPT 1993/09-836}
\centerline{hep-ph/9309334}
\centerline{September 1993}
\vskip 2truecm
\noindent
\hrule
* Contribution to the proceedings of the Workshop on Physics and Experiment at
Linear $e^+ e^-$ Colliders - Munich, Annecy, Hamburg 1993. 
Work supported in part by the Swiss National Foundation.
\hfill\break\noindent
a) Dipartimento di Fisica, Univ. di Firenze, I-50125 Firenze, Italy.
\hfill\break\noindent
b) I.N.F.N., Sezione di Firenze, I-50125 Firenze, Italy.
\hfill\break\noindent
c) CPT, CNRS, Luminy Case 907, F-13288, Marseille, France.
\hfill\break\noindent
d) D\'ept. de Phys. Th\'eor., Univ. de Gen\`eve, CH-1211 Gen\`eve 4.
\vfill
\eject
\null
$$\vbox{\vskip 1.5truecm}$$
\centerline
{\bf ABSTRACT}
\vskip 1truecm
\noindent
We consider the possibility of detecting vector resonances from a strong
electroweak sector, in the framework of the BESS model, at future 
$e^+ e^-$ colliders up to the TeV range. If the mass $M_V$ of the new vector 
boson multiplet is not far above or if it is below the maximum machine energy,
such a contribution would be manifest.
The process of $W$-pair production by $e^+e^-$ annihilation allows for 
sensitive tests of the strong sector, especially if the $W$ polarizations are 
reconstructed. 
\vfill
\eject
\hsize=16true cm
\vsize=23.5true cm
\overfullrule=0pt

\def\gp{{g^\prime}}
\def\gs{{g^{\prime\prime}}}

\def \e{{\rm e}}
\def \rs{\sqrt{s}}
\newcount \nfor

\def \form {\global \advance \nfor by 1 \eqno(1.\the\nfor)}
\noindent
{\bf 1. THE BESS MODEL}
\bigskip
We consider the sensitivity of $e^+e^-$ linear colliders, for different total 
center of mass energies and luminosities, to the BESS model which corresponds 
to a breaking of the electroweak symmetry due to a strongly interacting 
sector [1].
In BESS the electroweak symmetry breaking is obtained via a non-linear
realization and no Higgs particles are present. The nonlinear realization can
be seen classically as the limit of infinite Higgs mass for the corresponding
linear realization of the standard model. The effective lagrangian based, in 
its minimal version on the nonlinear breaking $SU(2)_L\otimes SU(2)_R
\otimes SU(2)_V \to SU(2)_D$, provides new gauge bosons $(V^\pm,V^0)$ that,
together with the hypothesis that quantum effects provide their kinetic
terms, appear as dynamical bound states of the strongly interacting sector. 
These new gauge bosons can be produced as real resonances if their mass is 
below the collider energy.
Because of beamstrahlung and synchrotron radiation, in a high energy collider,
one expects to see dominant peaks below the maximum c.m. energy 
even without having to tune the beam energies.
If instead the masses  of the $V$ bosons are  higher than the maximum c.m.
energy, they would give rise to indirect effects in the
$e^+e^-\rightarrow f^+f^-$ and  $e^+e^-\rightarrow W^+W^-$  cross sections,
due to their mixing with the electroweak gauge bosons and their fermion
couplings. Our description of the vector resonances provides a quite general 
background for testing the idea of a strong interacting electroweak sector 
and for example, after specialization of the parameters, it can describe a 
standard techni-$\rho$ state. 
\par
The BESS model contains as additional parameters the
mass $M_V$ of the new bosons which are a degenerate triplet of an additional
$SU(2)$ gauge group, their gauge coupling $\gs$ which
is assumed to be much larger than $g$ and $\gp$, and a parameter
$b$ specifying the direct coupling of $V$ to fermions. The standard model (SM) 
is formally recovered in the limit $\gs\to\infty$, and
$b=0$. Mixings of the ordinary gauge bosons to the $V$'s are of the order
$O(g/\gs)$. Due to these mixings, $V$ bosons are coupled to fermions
even for $b=0$. Furthermore these couplings are still present in
the $M_V\to\infty$ limit, and therefore the new gauge boson effects
do not decouple in the large mass limit.
\par
The precise measurements of the electroweak parameters done at LEP give 
the possibility of finding some hints for going beyond the standard model. 
Through mixing effects the contribution of vector resonances from the strong 
sector affects masses and couplings of ordinary gauge bosons. 
Therefore precise measurements of the width
of $Z$, its mass and forward backward asymmetries, performed at LEP,
allow for restrictions on the unknown parameters of the BESS model:
the mass $M_V$, the direct coupling to fermions $b$, and the gauge
coupling constant $\gs$ of the $V$ bosons.
\par
BESS is a non-renormalizable effective theory and when radiative corrections
are considered, a cut-off $\Lambda$ has to be introduced. This cut-off
plays the role of the parameter $m_H$, the Higgs mass of the SM.
We shall assume for BESS the same one-loop radiative corrections of the 
standard model interpreting $m_H$ as the cut-off.
\par
Using the following LEP data and the
CDF/UA2 measurement of the  mass ratio $M_W/M_Z$ [2]
$$
\eqalign{
&M_Z=91.187\pm 0.007~GeV\cr
&\Gamma_Z=2488\pm 7~MeV\cr
&\Gamma_h=1740\pm 6~MeV\cr
&\Gamma_{\ell}=83.52\pm 0.28~MeV\cr
&A_{FB}^{\ell}=0.0164\pm 0.0021\cr
&A_{\tau}^{pol}=0.142\pm 0.017\cr
&A_{FB}^{b}=0.098\pm 0.009\cr
&{M_W\over M_Z}=0.8798\pm 0.0028}
\form
$$
we obtain bounds on the BESS model, that we express as $90\%$ C.L.
contours in the plane $(b,g/\gs)$ for given $M_V$ (see Fig. 1).
The bounds depend mainly on the large range allowed at present for the SM 
parameters $m_{top}$ and $\alpha_s$. For this reason we show in Fig. 1  
the total allowed region for $m_{top}$ and $\alpha_s$ within the indicated 
ranges.
 
 
\baselineskip=10pt
\smallskip
\noindent
{\bf Fig. 1} - {\it  $90\%$ C.L. contour in the plane $(b,g/\gs)$
for $M_V=600~GeV$, from LEP and CDF/UA2 data
$(130\le m_{top}(GeV)\le 180$, $0.11\le\alpha_s\le 0.13$, and $\Lambda=1~TeV)$.}
\smallskip
 
\baselineskip=14pt
 
The bounds in Fig. 1 are almost independent of the mass of the new resonances 
$V$ and of the choice of the cut-off, while they become stronger for increasing 
$\alpha_s$ and  $m_{top}$. The latest data given at the Marseille Conference
[3] do not improve significantly the restrictions on the parameter space
of the BESS model shown in Fig.~1. LEP 200 is expected to increase only 
marginally the sensitivity over LEP. The relevant modification will be 
brought by more accurate determination of $M_W$.       
\par
In order to compare with a standard techni-$\rho$, one has to take
$b=0$, and
$$
{g\over\gs}=\sqrt{2}{M_W\over M_{\rho_T}}
\form
$$
and identify $M_{\rho_T}$ with $M_V$ [4].
\bigskip
\newcount \nfor

\def \form {\global \advance \nfor by 1 \eqno(2.\the\nfor)}
\noindent
{\bf 2. THE FERMIONIC CHANNEL}
\bigskip
The fermionic channel $e^+e^-\rightarrow f^+f^-$ at linear $e^+ e^-$ colliders 
improves the existing limits on BESS from LEP1 and UA2/CDF only for energies 
in the range $300-500~GeV$ and using longitudinally polarized electron beams.
Our analysis is based on the observables:
$$\eqalign{
&\sigma^{\mu},~~R=\sigma^h/\sigma^{\mu}\cr
&A_{FB}^{e^+e^- \to \mu^+ \mu^-},~~ A_{FB}^{e^+e^- \to {\bar b} b}\cr
&A_{LR}^{e^+e^- \to \mu^+  \mu^-},~~A_{LR}^{e^+e^- \to {\bar b} b},~~
A_{LR}^{e^+e^- \to {had}} \cr
}\form
$$
where
$A_{FB}$ and $A_{LR}$ are the forward-backward and left-right asymmetries,
and $\sigma^{h(\mu)}$ is the total hadronic (muonic) cross section.
The relevant formulas for this study have been reported in [5].
 
In the numerical analysis, following the existing studies of 500
GeV $e^+e^-$ linear
colliders [6], we assume a relative systematic error in luminosity of
${{\delta {\cal L}}/ {\cal L}}=1\%$ and ${{\delta\epsilon_{\rm hadr}}/
\epsilon_{\rm hadr}}=1\%$ (which is perhaps an optimistic choice due to the
problems connected with the $b$-jet reconstruction),
$\delta\epsilon_{\mu}/\epsilon_{\mu}=0.5\%$,
where $\epsilon_{hadr,\mu}$ denote the selection efficiencies. We shall
also assume the same systematic errors for the 1 and 2 $TeV$ machines.
Finally we have considered an integrated luminosity $L=20~fb^{-1}$. This 
integrated luminosity would correspond
to about one year ($10^7~sec.$) of running. One should of course take
into account beamstrahlung effects. However for two body final states,
as we consider here, the practical effect is a reduction of the luminosity.
This means that with the assumed nominal luminosity one has to run for a
correspondingly longer period.
\par
In the case the mass of the resonance could not be reached we can get 
restrictions on the parameter space by combining the observables of eq. (2.1). 
Throughout this paper we assume $m_{top}=150~GeV$ and $\Lambda=1~TeV$. Our 
results are shown in Fig. 2. Unfortunately the most sensitive observables
are the left-right asymmetries, which means that, if the beams are not
polarized, one gets practically no advantage over LEP1 from this channel.
\par
The contours shown in Fig. 2 correspond to the regions which are allowed
at 90\% C.L. in the plane $(b,g/\gs)$, assuming that the BESS deviations
for the observables of eq. (2.1) from the SM predictions are within the
experimental errors. The results are obtained assuming a longitudinal
polarization of the electron $P_e=0.5$ (solid line) and $P_e=0$ (dashed line).
We assume $\sqrt{s}=500~GeV$ and $M_V=600~GeV$.
 
 
\baselineskip=10pt
\smallskip
\noindent
{\bf Fig. 2} - {\it $90\%$ C.L. contours in the plane $(b,g/\gs)$ for $\sqrt 
s=500~GeV$ and $M_V=600~GeV$ from the fermionic channel (we assume 
$m_{top}=150~GeV$, $\alpha_s=0.12$ , $\Lambda=1000~GeV$). The solid line 
corresponds to polarization $P_e=0.5$ while the dashed line is for unpolarized
electron beams. The allowed regions are the internal ones.}
\smallskip
 
\baselineskip=14pt
 
As it is clear there is no big improvement
with respect to the already existing bounds from LEP1. Increasing the
energy of the machine does not drastically change the results.
We have also explored the sensitivity with respect to $M_V$, by choosing
$b=0$. We find that, even for polarized electron beams, the bounds improve 
only for $M_V$ close to the value of the energy of the machine.
\bigskip
\newcount \nfor

\def \form {\global \advance \nfor by 1 \eqno(3.\the\nfor)}
\noindent
{\bf 3. THE $WW$ CHANNEL}
\bigskip
In this section we consider the channel $e^+e^-\rightarrow W^+W^-$, 
which is expected to
be more sensitive, at high energy, than the $f\bar f$ channel to effects
coming from a strongly interacting electroweak symmetry breaking sector.
In the case of a vector resonance this is due to the strong coupling
between the longitudinal $W$ bosons and the resonance. Furthermore this
interaction, in general, destroys the fine cancellation among the $\gamma-Z$
exchange and the neutrino contribution occurring in the SM. This effect gives
rise, in the case of the BESS model, to a differential cross-section
increasing linearly with $s$. However one can show that the leading term in 
$s$ is suppressed by a factor $(g/\gs)^4$. Therefore the effective deviation 
at the energies considered here is given only by the constant term, which is 
of the order $(g/\gs)^2$.
\par
We consider one $W$  decaying leptonically and the other hadronically. The 
main reason is that in this way we get a clear signal useful to reconstruct the 
polarization of the $W$s [7]. Let us consider the observables:
$$
\eqalign{
&{d\sigma \over {d\cos\theta}}(e^+ e^-\to W^+ W^-)\cr
& A_{LR}^{{ e^+ e^- \to W^+ W^-}}=(
{d\sigma \over {d\cos\theta}}(P_{e}=+P)-
{d\sigma \over {d\cos\theta}}(P_{e}=-P))/
{d\sigma \over {d\cos\theta}}\cr}
\form
$$
where $\theta$ is the $e^+e^-$ center of mass scattering angle.
Assuming that the final $W$ polarization can be reconstructed by using the
$W$ decay distributions, we examine the cross sections for $W_LW_L$, 
$W_TW_L$, and $W_TW_T$. The relevant formulas have been given in [5].
In the case of the $WW$ channel we assume ${{\delta B}/ B}=0.005$ [8], where
$B$ denotes the product of the branching ratio for  $W\rightarrow hadrons$
and that for $W\rightarrow leptons$, and we assume $1\%$ for the acceptance.
\par
To discuss the restrictions on the parameter space for masses of the
resonance a little higher than the available c.m. energy we have taken into
account the experimental efficiency. We have assumed
an overall detection efficiency of 10\% including the branching ratio
$B=0.29$ and the loss of luminosity from
beamstrahlung [7]. This gives an effective branching ratio of about 0.1. 
\par
For a collider at
$\rs=500~GeV$ the results are illustrated in Fig. 3. The contours have
been obtained by taking 18 bins in the angular region restricted by
$|\cos\theta|< 0.95$. This figure illustrates the 90\% C.L. allowed regions
for $M_V=600~GeV$
obtained by considering the unpolarized $WW$ differential cross-section
(dotted line), the $W_LW_L$ cross section (dashed line),
and the combination of the left-right asymmetry with all the
differential cross-sections for the different final $W$ polarizations
(solid line). We see that already at the level of the
unpolarized cross-section we get important restrictions with respect to LEP1. 
 
 
\baselineskip=10pt
\smallskip
\noindent
{\bf Fig. 3} - {\it $90\%$ C.L. contours in the plane $(b,g/\gs)$ for $\sqrt 
s =500~GeV$ and $M_V=600~GeV$ from the unpolarized $WW$ differential cross 
section (dotted line), from  the $W_{L}W_{L}$ differential cross section
(dashed line) and  from all the differential cross sections for $W_{L}W_{L}$, 
$W_{T}W_{L}$, $W_{T}W_{T}$ combined with the $WW$ left-right asymmetries 
(solid line). The allowed regions are the internal ones.}
\smallskip
 
\baselineskip=14pt
\par
Fixing now $b=0$ we can see in Fig.~4 the restrictions in the plane
$(M_V, g/g")$ for three different choices of the collider energy, 
assuming an integrated luminosity of $20 fb^{-1}$. 
 
 
\baselineskip=10pt
\smallskip
\noindent
{\bf Fig. 4} - {\it $90\%$ C.L. contours in the plane $(M_V,g/\gs)$ for 
$\sqrt s=0.3,~0.5,~1~TeV$, $L=20~fb^{-1}$ and $b=0$. The solid line 
corresponds to the bound from the unpolarized $WW$ differential cross section, 
the dashed line to the bound from all the polarized differential cross 
sections $W_{L}W_{L}$, $W_{T}W_{L}$, $W_{T}W_{T}$ combined with the $WW$ 
left-right asymmetries. The lines give the upper bounds on $g/\gs$.}
 
\smallskip
\baselineskip=14pt
In Fig. 5 the upper bounds on $g/\gs$ for $M_V=1.5~TeV$ and $b=0$ are shown 
as a function of the center of mass energy of the $e^+ e^-$ collider in the 
case no deviation from the SM is found. The relevance of final W polarization 
reconstruction over the unpolarized cross section (solid line) is apparent 
in the whole energy range considered. A higher luminosity option is shown for 
the case $\sqrt{s}=1~TeV$ and $L=80~fb^{-1}$ (black dots).
 
 
\baselineskip=10pt
\smallskip
\noindent
{\bf Fig. 5} - {\it  $90\%$ C.L. contours in the plane $(\sqrt{s},g/\gs)$ for 
$M_V=1.5~TeV$, $b=0$ and $L=20~fb^{-1}$. The lines correspond to the 
unpolarized $WW$ differential cross section (solid line), the $W_{L}W_{L}$ 
differential cross section (dashed line), and all the differential cross 
sections for $W_{L}W_{L}$, $W_{T}W_{L}$, $W_{T}W_{T}$ combined with the $WW$ 
left-right asymmetries (dotted line) and from all the WW and fermionic
observables with $P_e=0.5$ (dash-dotted line) and represent the upper bounds 
on $g/\gs$. The black dots are the bounds for the unpolarized $WW$ 
differential cross section and from all the WW and fermionic observables at
$\sqrt{s}=1~TeV$ and $L=80~fb^{-1}$.}
\smallskip
\baselineskip=14pt
 
\bigskip
\newcount \nfor

\def \form {\global \advance \nfor by 1 \eqno(4.\the\nfor)}
\noindent
{\bf 4. FUSION PROCESSES}
\bigskip
$W^+ W^- $ pairs can also be produced through the fusion of a pair of ordinary 
gauge bosons, each of them emitted from an electron or a positron. In the 
effective-W approximation the initial $W,Z,\gamma$ are assumed to be real 
particles and the cross section for producing a  $W^+ W^- $ pair is obtained 
by means of a convolution of the fusion subprocess with the luminosities of
the initial $W,Z,\gamma$ inside electrons and positrons.
There are two fusion subprocesses which contribute to produce $W^+ W^- $ pairs.
The first one is $e^+e^- \rightarrow W^+_{L,T} W^-_{L,T} e^+ e^-$. It is 
mediated by $W^{\pm}$ and $V^{\pm}$ exchanges in the $t$ and $u$ channels.
The second fusion subprocess we consider is $e^+e^- \rightarrow W^+_{L,T} 
W^-_{L,T}{\overline \nu} \nu$. It is mediated by $\gamma, Z$ and $V^{0}$ 
exchanges in the $s$ and $t$ channels. We have included in the computation
contributions both from the gauge boson trilinear and quadrilinear couplings.
\par
The fusion processes may be relevant because they allow
to study a wide range of masses for the $V$ resonance from one
given $e^+e^-$ c.m. energy. In the $e^+e^-$ center-of-mass frame the 
invariant mass distribution $d\sigma/dM_{WW}$ is 
$$
\eqalign{
{{d\sigma}\over {dM_{WW}}}&={1\over 4\pi s} {1\over M^2_{WW}}
\sum_{i,j}\sum_{l1,l2}
\int^{M_{WW}^2/4}_{(p_T^2)_{min}}
d{p_T^2}\int^{-\log{\sqrt{\tau}}}_{\log{\sqrt{\tau}}}dy ~f^{l1}_i(\sqrt{\tau}
\e^{y})f^{l2}_j(\sqrt{\tau}\e^{-y})\cr
&\cdot {p'\over p} {1\over\sqrt{M^2_{WW}-4 p_T^2}}
|M({V^{l1}_i V^{l2}_j} \rightarrow W^+_{l3}
W^-_{l4})|^2}
\form
$$
where $p_T$ is the transverse momentum of the outgoing $W$,
$\tau=M_{WW}^2/s$,
$p$ and $p'$ are the absolute values of the three momenta
for incoming and outgoing pairs
of vector bosons: $p=(E_1^2-M_1^2)^{1/2}=(E_2^2-M_2^2)^{1/2}$ and
$p'=(\sqrt{M_{WW}}/2) (1-4 M_W^2/M_{WW}^2)^{1/2}$
where $E_i$ the fraction of the electron (or positron)
energy carried by the vector boson $V_i$ with mass $M_i$ and helicity $l_i$.
The structure functions $f$ appearing in formula (4.1) are given by:
$$
\eqalign{
f^+(x)=&{{\alpha_{em}}\over{4\pi}} {{[(v+a)^2+(1-x)^2(v-a)^2]}\over{x}}
\log{s\over{M^2}}\cr
f^-(x)=&{{\alpha_{em}}\over{4\pi}} {{[(v-a)^2+(1-x)^2(v+a)^2]}\over{x}}
\log{s\over{M^2}}\cr
f^{0}(x)=&{{\alpha_{em}}\over{\pi}} (v^2+a^2) {{1-x}\over{x}}
}
\form
$$
and represent the probability of finding inside the electron a vector boson of
mass $M$ with fraction $x$ of the electron energy. In eq. (4.2) $v$ and $a$
are the vector and axial-vector couplings of the gauge bosons to fermions.
For the detailed formulas for amplitudes and couplings see [5].
We do not find significant differences between the SM and the BESS model 
differential cross section in the case of the process $e^+e^-
\rightarrow W^+W^-e^+e^-$. This is due, on one hand, to the absence of the 
$s$ channel exchange of the $V$ resonance and on the other hand,
to the dominance of the $\gamma\gamma$ fusion  contribution, and to the
the fact that in BESS the couplings of the photon to the fermions and to 
$W^+W^-$ are the same as those of the SM.
\par
Let us now consider the process $e^+e^-\rightarrow W^+W^-\nu {\overline \nu}$. 
We have computed the differential cross sections $d\sigma/d M_{WW}$ both for 
the SM with $M_H=100~GeV$ and for the BESS model. The only interesting channel 
is the one corresponding to longitudinally polarized final $W$s.
The results are illustrated in Fig. 6 where we compare $d\sigma/\d M_{WW} 
(LL)$ for the SM (dashed line) and for the BESS model (solid line) for 
$\sqrt {s}=1.5~TeV$, $b=0.01$, $\gs=13$ and $M_V=1~TeV$. We apply only a cut 
for $(p_T)_{min}=10~GeV$.
 
 
\baselineskip=10pt
{\vskip 2truecm
\noindent
{\bf Fig. 6} - {\it Longitudinally polarized differential cross-section
$d\sigma/d M_{WW} (e^+ e^- \to W^+_L W^-_L \nu\bar\nu)$ (in $fb/GeV$) versus 
$M_{WW}$ for the SM (dash line) and BESS model (solid line) corresponding to 
$\sqrt s =1500~GeV$, $M_V=1000~GeV$, $b=0.01$, and $\gs=13$.}
}
\smallskip
 
\baselineskip=14pt
 
Integrating the differential cross section for $500<M_{WW}(GeV)<1500$
and even considering a high integrated luminosity of 80 $fb^{-1}$ we obtain
127 $W_L$ pairs for the SM and 158 for the BESS model (with $M_V=1~TeV$,
$\gs=13$ and $b=0.01$) corresponding to a statistical significance of 
only 2.75. This result is quite discouraging as we have still not
included the branching ratio. The situation does not improve significantly 
even when considering a wider resonance, varying the BESS parameters, or
considering $\sqrt{s}=2\;TeV$ in the region allowed by the present bounds 
(see Fig. 1).
\par
The fusion process in the charged channel $e^+e^- \rightarrow W^+_{L,T} 
Z_{L,T}{\overline \nu} e^-$ is even less encouraging.
In this case in fact, the SM cross section is bigger, being dominated by the
$\gamma W \rightarrow W Z$ fusion process, while the BESS effect
$ W Z \rightarrow V \rightarrow W Z$ is of the same order of magnitude and we 
expect a worse signal to background ratio.
\bigskip
\vfil
\eject
\newcount \nfor

\def \form {\global \advance \nfor by 1 \eqno(5.\the\nfor)}
\noindent
{\bf 5. CONCLUSIONS}
\bigskip
The result of our study is that the annihilation channels are by far the most
important ones in order to distinguish between the SM and the strong
electroweak sector as described through the BESS model at $e^+e^-$ colliders
for the considered energy range. 
In particular the process of $W$-pair production by $e^+e^-$ annihilation 
allows for sensitive tests of the strong sector, especially if the $W$ 
polarizations are reconstructed. Fusion processes may become more relevant at
higher energies or luminosities, but are still of minor interest at the 
energies and luminosities considered in this work. 
\par
The study performed for the case of a $e^+ e^-$ accelerator is complementary 
to those performed for $pp$ colliders. In fact proton colliders offer the 
possibility of studying the $V^\pm$ resonances through the $W^\pm Z$ decay [9], 
while the $V^0\rightarrow W^+W^-$ channel  is difficult to study due to 
background problems. On the contrary $e^+e^-$ colliders give the possibility 
of detecting new neutral vector bosons. This can be relevant in order to 
distinguish among BESS and other models. 
\par
Even in the case of a mass of the $V^0$ resonance higher than the c.m. energy 
of the collider, the process of $W$ pair production will allow to restrict the 
parameter space of BESS, especially if the $W$ polarizations can be
reconstructed.
\par
Already at $\sqrt{s}=500~GeV$ and integrated luminosity $L=20~fb^{-1}$, if 
no deviation from the SM prediction is found, the BESS model parameters $\gs$ 
and $b$ can be severely restricted. 
If higher energy colliders are available ($\sqrt{s}=1,2\; TeV$ and $L=20,80
\;fb^{-1}$), it is possible to get an upper bound on $g/\gs$ of the order of 
0.02 at $b=0$, for any given value of $M_V$.
\bigskip
\bigskip
\newcount \nref

\def\ref {\global \advance \nref by 1 \ifnum\nref<10 \item {$ [\the\nref]~$}
\else \item{$[\the\nref]~$} \fi}
\centerline{\bf REFERENCES}

\noindent
\ref  
R. Casalbuoni, S. De Curtis, D. Dominici and R. Gatto, Phys. Lett.
{\bf B155} (1985) 95;
Nucl. Phys. {\bf B282} (1987) 235.
\ref  
CDF Coll., F. Abe et al., Phys. Rev. Lett. {\bf 65} (1990) 2343;
UA2 Coll., J. Alitti et al., Phys. Lett. {\bf B276} (1992) 354;
C. De Clercq, to appear on the proceedings of the XXVIII Rencontres de
Moriond on Electroweak Interactions, Les Arcs, March 1993;
V. Innocente, {\it ibidem};
R. Tenchini, {\it ibidem};
G. Altarelli, talk given at  ``FILEP - Incontro sulla Fisica a LEP",
Firenze April 1-2 1993.
\ref  
G. Altarelli and P. Lefrancois, talks given at the International
Europhysics Conference on High Energy Physics, Marseille July 22-28
1993.
\ref  
R. Casalbuoni, S. De Curtis, D. Dominici, F. Feruglio and R. Gatto,
Phys. Lett. {\bf B269} (1991) 361.
\ref  
R. Casalbuoni, P. Chiappetta, A. Deandrea, S. De Curtis, D. Dominici, 
and R.~Gatto, Zeit. Phys. {\bf C60} (1993) 315.
\ref 
R. Casalbuoni, P. Chiappetta, S. De Curtis, D. Dominici, F. Feruglio
and R. Gatto, in "$e^+e^-$ Collisions at 500 $GeV$: the Physics Potential",
Proceedings of the Workshop, edited by P.M. Zerwas, DESY 92, 123B,
August 1992, p. 513;
A. Djouadi, A.~Leike, T. Riemann, D. Schaile, and C. Verzegnassi,
{\it ibidem} p. 491;
D. Dominici, in "Physics and Experiments with Linear Colliders",
Saariselk\"a, Finland, September 9-14, 1991, edited by R. Orava, P. Eerola
and M. Nordberd, World Scientific, p. 509.
\ref  
K. Fujii, KEK preprint 92-31, to appear in the Proceedings of
the 2nd KEK Topical
Conference on $e^+e^-$ Collision Physics,
KEK, Tsukuba, Japan, November 26-29 1991.
\ref   
M. Frank, P. M\"attig, R. Settles and W. Zeuner, in
"$e^+e^-$ Collisions at 500 $GeV$: the Physics Potential", Proceedings
of the Workshop, edited by P.M. Zerwas, DESY 92, 123A, August 1992,
p. 223.
\ref  
R. Casalbuoni, P. Chiappetta, S. De Curtis, F. Feruglio, R. Gatto,
B. Mele and J.~Terron, Phys. Lett. {\bf B249} (1990) 130, and
in "Large Hadron Collider Workshop" Proceedings of the Workshop, edited by
G. Jarlskog and D. Rein, p.786.
\vfill
\eject
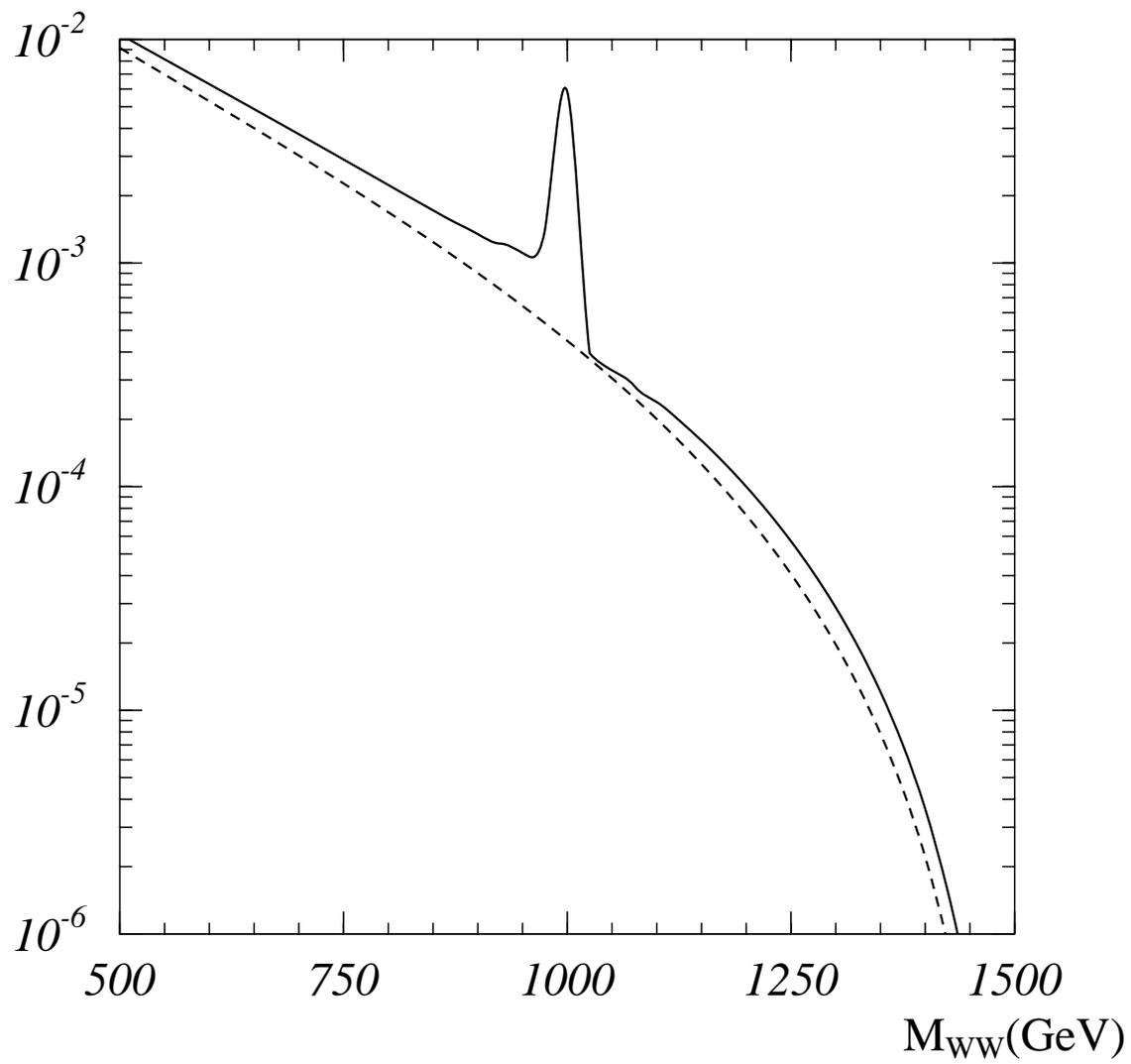
\bye